# Towards Energy Harvesting Powered Sensor Networks for Cyber-Physical Systems

Dr. Peng Hu

CMC Microsystems, Kingston, Canada

## Abstract

The concept of sensor networks (WSNs) has become an important component of the recently proposed cyber-physical systems (CPSs) and Internet of Things which can connect the physical world with embedded software systems. Energy-harvesting (EH) as an enabling technology applied in SNs can heavily reduce the installation and maintenance cost as well as increase system life, flexibility, and scalability as the EH is subject to the important "energy interruption" problem where system tasks can be significantly disrupted by the EH outputs. This problem not only affects the system modelling but also relates to the cross-layer network design and schedulability. This paper discusses the EH-powered CPS architecture and some theoretical problems. In particular, we propose a framework for fully EH-powered CPS, within which essential topics are discussed, such as EH software architecture, EH models, and real-time communication. Different EH models for CPSs (e.g., fully EH-powered and partially EH-powered) are addressed. Future research directions are briefly discussed in the end.

## 1 Introduction

The concept of sensor networks (SNs) has become an important component of the recently proposed cyber-physical systems (CPSs) and Internet of Things with which can connect the physical world and software systems. In each of these system paradigms, the autonomous system is an main design goal where a fair amount of autonomy can be achieved by power modules which nowadays can be zero-battery or self-powered. With the increasing capacity of EH modules and decreasing power consumption of embedded systems modules, zero-battery sensor networks enabled by energy harvesting (EH) modules have become more and more popular as they can heavily reduce the installation and maintenance cost as well as increase system life, flexibility, and scalability [1]. In real-word examples, the EH-powered sensor networks have been applied in industrial process monitoring and consumer applications such as healthcare monitoring with body area sensor networks. Because EH-based SNs is subject to "energy interruptions" when EH techniques are employed, the EH-powered system design is quite different from the traditional SNs. The energy interruptions not only affect the system modelling but also relate to the cross-layer network design and schedulability.

When it comes to CPS, an important problem of the EH-powered CPS is whether the EH module can supply enough power to task executions. This problem relates to the software design and thereafter the software design topics in firmware, real-time software, and communication protocols should be considered.

In this paper, we focus on the discussion about the software framework based on the the nano-/micro-technology based energy harvesters. We will discuss the theoretical frameworks and the system architecture supporting power-level abstraction, real-time communication and task scheduling. Different EH models for CPSs will be defined and discussed. We summarize the discussion and propose the future research topics in the end.

## 2 Related Work

The advancement of RF modules and embedded hardware has been driving the low-power SN applications and opens the opportunity of employing the EH modules in the CPS applications. In this sense, the EH-powered CPS can initially be split into two aspects, i.e., EH hardware and EH software. The EH hardware modules provide the basic power profile and capacity of the system while the EH software modules provide the functions of utilizing the power budget to meet the application-specific goals. The EH hardware-software co-design is a way to build an EH-powered CPS but it currently depends on the application-specific system design requirements where some insights into the common system design can deeply strengthen the general CPS system performance.

The current EH technologies include different kinds of technologies that can provide power in a range from several microwatts to few miniwatts. The sources of the EH modules include stress-strains, vibration, thermal gradient, infrared, RF, and biochemistry. An energy tree was designed [2] to utilize wind energy with leaf-sized piezo-electric generators. The newly proposed WiTricity solution [3, 4] can be considered as an RF-based EH technology, where the wirelessly powered communication devices can provide sufficient power outputs that meet many wireless data transmissions and enable a generation of new wireless communication models. A novel monitoring application using microprocessor thermal gradient with thermogenerators was introduced in [5]. However, so far the important theoretical questions such as heterogeneity of EH outputs and scalable EH framework with EH module software which can take multiple EH sources have not been systematically addressed.

In typical SN applications, communication is the major factor of system power consumption. This factor results in that some discussions in the literature associate the data rate metrics in wireless data communications with EH technologies. However, considering the Moore's law of microelectronics industry, we do not limit the discussion of the current EH technologies as the current technological trends indicate that the transceiver power consumption and EH related technologies can finally come to a good match, which, in other words, means more and more SN applications with different data rate requirements can be fully EH powered. The specific EH models have not been addressed in the literature. Therefore, we will start to address general EH models that can be used in the EH-powered CPS.

## 3  System Architecture

An EH-powered system can be considered to have the architecture shown in Figure 1, where a CPS device is seen as a typical embedded system with an EH module. Figure 1(a) shows the general system architecture of an EH-powered sensor node where EH is associated with all key components of the system. In Figure 1(a), the EH management software and EH adaptation interface provide essential EH services to the traditional embedded SN components, while the latter provides an abstraction interface to the embedded operating system (OS) which can make sure the successful task scheduling and interruption operations. Although it is possible to integrate the EH SW into the embedded OS, there are two reasons we do not put the EH SW on top of the OS: 1) it can cause overhead for the EH component and the energy-dependent tasks; and 2) EH components are various and they need efficient different drivers from EH vendors. As such, EH adaptation can provide an interface to the OS to invoke the essential EH modules. In addition, the external EH can provide a smooth integration to the existing OS.

To see how the EH adaptation interface works, we can consider an example that when the embedded OS tries to schedule the data transmission task, the EH adaption interface will interact with the EH management software module and get the temporal EH power availability information. In Figure 1(b), the detailed components in energy harvesting module are shown, where the EH management software interact with heterogeneous EH source modules with subcomponents including EH status management, EH indicator, EH switcher, and EH task scheduler. The EH status management component is responsible of the status control and monitor of the EH sources as well as response to the commands from external entities, while the EH indicator component provides EH power information to the external EH adaptation interface, such as EH power profile, temporal energy availability, and EH source being used. The EH switcher component deals with the switching among the EH sources in order to meet the requirement of the internal EH task scheduler. The EH task scheduler component tries to provide the guaranteed power output and make optimal scheduling of hybrid EH sources. In summary, the key sub components shown in Figure 1(b) are expected to deal with the general interactions from external embedded systems applications with a common interface.

### 3.1  Energy Harvesting Model

We can categorize the current EH-powered CPS into three types: fully EH-powered, partially EH-powered, and non-EH-powered. Non-EH-

powered model will not be discussed as it is the most common battery-powered model. The other two models are defined in the following.

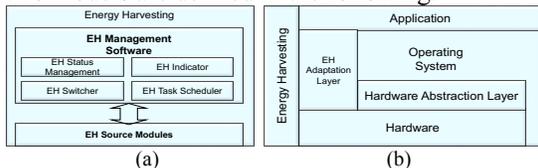

Figure 1: (a) System architecture of a CPS device, and (b) essential energy harvesting components

**Definition 1**(Fully EH-powered CPS): a CPS with power supply provided fully by EH modules for completing desired functions of a CPS for a period of time is said to be a fully EH-powered (FEHP) CPS.

There are four possible scenarios associated with the FEHP model: a) the EH power can be converted into stable power outputs that can meet task schedulability requirements; b) the EH power can be converted into stable power outputs that cannot meet task schedulability; c) the EH power can be converted into power outputs in variable duty cycles which task schedulability need to consider these temporal characteristics; and d) the EH power cannot provide usable power outputs for required tasks.

**Definition 2**(Partially EH-powered CPS): a CPS with power supply provided partially by EH modules for completing desired functions of a CPS for a period of time is said to be a partially EH-powered (PEHP) CPS.

In this PEHP model, the CPS can be powered by hybrid power sources including batteries, and therefore the stable power output is likely to be available for some tasks. As such, in this paper, we focus on the discussion of the FEHP model.

## 3.2 EH-Powered Tasks

With the FEHP model, the CPS tasks are only EH-powered which result in the fundamental difference from the traditional SNs in terms of the task execution and schedulability.

We first show this difference in Figure 2, where in the traditional non-EH-powered CPS, the power supply module does not have interactions with sensor nodes or sensors, while in the EH-powered CPS, the EH module has direct interactions with sensor nodes and indirectly coupled interactions with the sensors. Specifically, in Figure 2(b), when the gateway G initiates a sensing information retrieval task, the sensor node S responds to the request from G by obtaining the sensing information from attached two sensors T and A; however, this task cannot be completed without E as E provides temporal energy information for sensor which can be obtained by the EH management software interface. If we denote the link between S and T as $L_{SG}$ and denote the energy availability link at time $t_i$ of $L_{SG}$ as $L_{SG}(t_i, t_{i-1})$ where $\Delta t = t_i - t_{i-1}$ is the period of time that the required power supply for the task is available. Then the $L_{SG}(t_i, t_{i-1})$, $L_{SA}(t_i, t_{i-1})$, $L_{SE}(t_i, t_{i-1})$ are the actual link we can consider during the task execution. In this way, the task schedulability is shown to be heavily dependent on the EH power outputs from E, while the case of Figure 2(a) does neither have such temporal property of links nor the task schedulability dependency on E. Furthermore, it is not difficult to see that with the increasing number of sensors in addition to T and A as well as the mesh networking with multiple sensor nodes as intermediate routers, the scheduling will be more complicated.

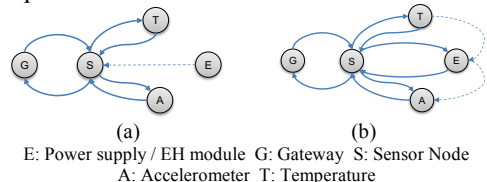

E: Power supply / EH module  G: Gateway  S: Sensor Node
A: Accelerometer  T: Temperature

Figure 2: Example task state flow of the (a) traditional CPS and (b) EH-powered CPS. The solid arrows denote the direct relationship and the dotted arrows denote the indirect relationship during a task execution.

## 3.3 EH Output Patterns

We need to discuss the EH output patterns as its property directly relates to the task schedulability. Among different energy output patterns, we consider the commonly used periodic sawtooth wave pattern [5]. Then, we show that the task schedulability issue can be mitigated by using additional EH sources. In Figure 3 (a), the voltage output of a single EH source is shown against time, where the resultant square waves with period $T_1$ (indicated by shaded areas in Figure 3(a)) are the regulated outputs for the CPS. Note that in this case the EH-powered tasks tend to be split into time slots with length $T_1$. In Figure 3(b), two EH sources with two kinds of periodic sawtooth waves are shown, and there are two types of resultant square waves with periods $T_1$ and $T_2$ accordingly. In this case, the resultant period $T_{12}$ with the existence of the two EH sources has the bound [max($T_1,T_2$), $T_1+T_2$]. This result can be generalized to the case of a CPS having $m$ EH sources greater than two with periodic voltage outputs, where the resultant period $T$ has the bound [max($T_1$, $T_2$, …, $T_m$), $\Sigma T_i$]. For aperiodic cases, the results need to be modified with more information. For example, for the aperiodic resultant duty cycles following an on-off stochastic process, the upper and lower bound of $T$ has a more complex form characterized by the probability density functions.

After we quantitatively represent the resultant period of multiple or possibly heterogeneous EH sources, we need to think about how a task schedulability problem can take advantage of this. As an example shown in Figure 3, the time division multiple access (TDMA)-based scheduling would be a good fit to the communication task scheduling in the EH-powered CPS. Moreover, we should note that what we have discussed is the general FEHP solution where no advanced power storage devices are used. If power storage devices are added to the EH module, its outputs are expected to be more stable and consistent characterized by a longer expected value of the EH output period E[$T$].

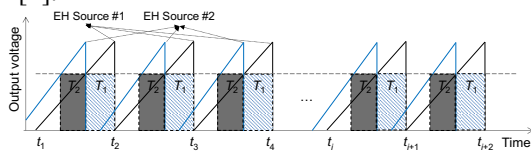

Figure 3: Sawtooth waves outputs and power availability (shown as square shaded areas) of two EH sources.

## 3.4 Real-Time Communication

The commonly used contention-based and TDMA-based protocols in SNs are two candidates for real-time communication protocols, where contention-based approach even with control of rate, contention window, back-off mechanism, or admission cannot provide predictable packet latencies and high throughput capacity. As such, TDMA-based real-time communication protocols are considered. Here our discussion is based on the IEEE 802.15.4 based superframe duration scheduling (SDS) scheme [6], which is shown in Figure 4. In the two-hop coordinator scenario, the superframe is defined by the so-called personal area network (PAN) coordinator; however, the introduction of EH outputs (which are assumed to be the same and synchronized for each coordinator) affects this superframe structure, where in the case shown in Figure 4 the active timeframe of coordinators is shortened by $\Delta t_{b1}$ and the beacon timeframe is shorten by $\Delta t_{b2}$. It is straightforward to see in Figure 4 that the periodic EH outputs in multi-hop SN affect the original SDS scheme.

One solution to this issue is to align superframe and beacon frame with the duty cycle of the EH outputs. In the case of Figure 4, the beacon frame has the length $\Delta t_{b1} + \Delta t_{b2}$ while the superframe has the length of EH resultant period $T$. If aperiodic EH outputs are considered in Figure 4, the EH outputs can be generalized as on-off process where on-periods with "Active" state shown in Figure 4 and off-periods are i.i.d. positive random variables. Therefore, the EH-powered CPS is like an opportunistic contention-based mechanism but the predictability of packet scheduling can still be achieved given a EH output process.

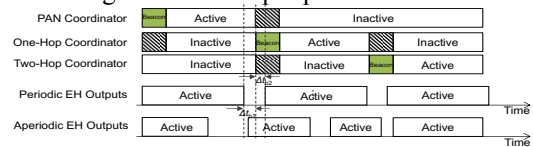

Figure 4: Superframe duration scheduling with periodic and aperiodic EH outputs

## 4  Conclusions

The EH-powered SN is an important paradigm of the CPS and Internet of Things, and its software architecture is an important topic yet to be explored. In this work, we open the discussion of some important topics in EH-powered CPS, such as the EH embedded system architecture, EH model, EH-powered tasks, and task schedulability. This paper also proposes the EH software architecture that can inspire further studies. There are several challenges in this proposed architecture, such as heterogeneous EH source integration, EH software verification and monitoring approach, energy intelligence/prediction algorithm, and interface standard.